\documentclass{article}
\usepackage{spconf,amsmath,graphicx}
\usepackage{amsfonts}
\usepackage{amssymb} 
\usepackage{color}
\usepackage{balance}

\usepackage{enumitem}

\usepackage{algorithm}
\usepackage{algpseudocode}
\usepackage{amsthm} 
\usepackage{graphicx}
\usepackage{multicol}
\usepackage{cite}

\usepackage[colorlinks=true,allcolors=black]{hyperref}

\newtheoremstyle{defstyle} 
    {0pt}                    
    {0pt}                    
    {\normalfont}            
    {}                       
    {\bfseries\itshape}      
    {.}                      
    {.5em}                   
    {\thmname{#1} \thmnumber{#2} \textbf{\thmnote{(#3)}}}    

\theoremstyle{defstyle}

\setlist[enumerate,1]{nosep, leftmargin=*}
\setlist[itemize,1]{nosep, leftmargin=*}

\def\x{{\mathbf x}}
\def\y{{\mathbf y}}

\def\X{{\mathbf X}}

\def\T{{\mathbf T}}

\def\S{{\mathbf S}}

\def\W{{\mathbf W}}
\def\B{{\mathbf B}}

\def\n{{\mathbf n}}

\title{Topological Neural Networks over the Air\vspace{.2cm}}

\name{Simone Fiorellino, Claudio Battiloro, and Paolo Di Lorenzo}
\address{\thanks{Fiorellino is with DIAG department, Sapienza University of Rome, via Ariosto 25, Rome, Italy. Battiloro and Di Lorenzo are with DIET Department, Sapienza University of Rome, Via Eudossiana 18, Rome, Italy. Di Lorenzo is also with Consorzio Nazionale Interuniversitario per le Telecomunicazioni (CNIT), Parma, Italy. {E-mail: \{simone.fiorellino, claudio.battiloro, paolo.dilorenzo\}@uniroma1.it}. This work was funded by the European Union under the Italian National Recovery and Resilience Plan (NRRP) of NextGenerationEU, partnership on “Telecommunications of the Future” (PE00000001 - program “RESTART”), and by the 6G-GOALS project under the 6G SNS-JU Horizon program, n.101139232.}\vspace{-1cm}}
\begin{document}
\ninept
\maketitle

\begin{abstract}
Topological neural networks (TNNs) are information processing architectures that model representations from data lying over topological spaces (e.g., simplicial or cell complexes) and allow for decentralized implementation through localized communications over different neighborhoods. Existing TNN architectures have not yet been considered in realistic communication scenarios, where channel effects typically introduce disturbances such as fading and noise. This paper aims to propose a novel TNN design, operating on regular cell complexes, that performs over-the-air computation, incorporating the wireless communication model into its architecture. Specifically, during training and inference, the proposed method considers channel impairments such as fading and noise in the topological convolutional filtering operation, which takes place over different signal orders and neighborhoods. Numerical results illustrate the architecture's robustness to channel impairments during testing and the superior performance with respect to existing architectures, which are either communication-agnostic or graph-based.

\end{abstract}
\begin{keywords}
Topological Signal Processing, Topological Neural Networks, Over-the-Air Computation, Cell Complexes. 
\end{keywords}

\section{Introduction}
\label{sec:intro}

In the last few years, there has been a large interest in developing methodologies to analyze, process, and learn from data defined over the vertices of a graph \cite{shuman2013emerging,ortega2018graph}. To this aim, several processing tools have been designed hinging on different graph shift operators (e.g., adjacency, Laplacian, etc.), thus leading to key contributions such as graph Fourier analysis, convolutional filtering, topology inference, and so on \cite{ortega2018graph}. 
However, despite their overwhelming popularity, graph representations can only take into account pairwise relationships among data. In several complex interconnected systems, relationships among data cannot be reduced to be only pairwise since multi-way interactions naturally occur among multiple entities. As an example, in gene regulatory networks, multi-way links among complex substances (i.e., genes) exist \cite{lambiotte2019networks}; also, in social networks, individuals typically join groups that clearly have a multi-way connectivity pattern. To incorporate multi-way relationships and develop suitable processing tools, we need to go beyond graphs, exploiting topological spaces having a rigorous algebraic description such as, e.g., simplicial complexes, cell complexes, cellular sheaves, and others \cite{grady2010discrete}. These considerations have sparked a strong interest in developing processing tools taking into account multi-way relationships among data, thus leading to the emergent field of topological signal processing (TSP) \cite{barbarossa2020topological,schaub2021signal}. In this context, the seminal works in \cite{barbarossa2020topological,schaub2021signal} illustrated the benefits obtained by processing signals defined over simplicial complexes. Then, the work in \cite{yang2022simplicial} proposed FIR filters for signals defined over simplicial complexes, hinging on a Hodge decomposition and higher-order combinatorial Laplacians.

Driven by the success of TSP, several deep neural network architectures able to learn from data defined over specific topological spaces (e.g., simplicial or cell complexes) have been developed, see, e.g., \cite{bodnar2021weisfeiler,ebli2020simplicial,roddenberry2019hodgenet,giusti2021san,effsimprep2022yang,battiloro2022tangent,battiloro2023tangentlearn, hajij2023topological, giusti2023cell}. Although Topological Neural Networks (TNN) have achieved remarkable results, they often operate under idealized conditions, e.g., considering perfect communication scenarios. This becomes a significant limitation in real-world applications where wireless communication typically introduces several impairments, such as fading and noise, thus affecting the transmission quality \cite{belloni2004fading, peppas2011overview,popa2008fading}. In such environments, each network agent receives faded/noisy messages from its neighbors, leading to corrupted outputs mismatching the one assuming ideal communication at train time, thus resulting in performance degradation at inference time. To cope with this issue, several works have been proposed for deep neural network training over wireless channels, see, e.g., \cite{o2016unsupervised,farsad2017detection,dorner2017deep,o2018over,bourtsoulatze2019deep,Jankowski2021AirNet,airgnns}. Specifically, the work in \cite{dorner2017deep} proposed end-to-end communication systems based on deep learning. Also, the work in \cite{Jankowski2021AirNet} shows that incorporating channel noise at train time can make deep neural networks more robust when network parameters are delivered over noisy communication channels. Then, the work in \cite{airgnns} proposes a variation of graph neural networks, named AirGNN, which considers channel fading and noise when aggregating features from neighbors, thus improving the architecture robustness to channel impairments in the inference phase. However, to the best of our knowledge, no previous works investigated TNNs over realistic wireless channels.

\noindent\textbf{Contribution.} In this paper, we introduce Topological Neural Networks over-the-air (AirTNNs), a novel framework that seamlessly integrates wireless communication channels within TNNs operating over regular cell complexes. Specifically, an AirTNN is a multi-layered architecture built by cascading novel cell complex filters over the air and pointwise nonlinearities. Cell complex filters over the air are based on shifting topological signals over wireless communication channels in an uncoded manner, capturing information from the multi-hop lower and upper neighborhoods induced by the complex. AirTNNs find natural application in scenarios where it is of interest to learn from topological data (e.g., signals defined over edges and/or polygons), which are collected by sensors that communicate over wireless channels to enable distributed processing. Several examples can be found in applications related to the monitoring and control of critical network infrastructures, such as traffic, hydraulic, or communication networks. Finally, we evaluate the effectiveness of AirTNNs on a source localization task, showing that they outperform both AirGNNs and baseline GNNs and TNNs.

\vspace{-.2cm}
\section{Background}
\vspace{-.1cm}
This section reviews some useful basics of TSP over cell complexes.

\noindent \textbf{Regular Cell Complex.}
    A {\it regular cell complex}  is a topological space $\mathcal{X}$ having a partition $\{\mathcal{X}_{\sigma}\}_{\sigma \in \mathcal{P}_{\mathcal{X}}}$ of subspaces $\mathcal{X}_{\sigma}$ of $\mathcal{X}$ called \textit{cells}, where $\mathcal{P}_{\mathcal{X}}$ is the indexing set of $\mathcal{X}$, such that \cite{grady2010discrete}:
    \begin{enumerate}
        \item For each $c\in\mathcal{X}$, every sufficient small neighborhood of $c$ intersects finitely many $\mathcal{X}_{\sigma}$;  
        \item For all $\tau$ and $\sigma$ we have that $\mathcal{X}_{\tau}$ $\cap$ $\overline{\mathcal{X}}_{\sigma}$ $\neq$ $\varnothing$ iff $\mathcal{X}_{\tau}$ $\subseteq$ $\overline{\mathcal{X}}_{\sigma}$, where $\overline{\mathcal{X}}_{\sigma}$ is the closure of the cell;\label{cond.2}
        \item Every $\mathcal{X}_{\sigma}$ is homeomorphic to $\mathbb{R}^{k}$ for some $k$;
        \item For every $\sigma$ $\in$ $\mathcal{P}_{\mathcal{X}}$ there is a homeomorphism $\phi$ of a closed ball in $\mathbb{R}^{k}$ to $\overline{\mathcal{X}}_{\sigma}$ such that the restriction of $\phi$ to the interior of the ball is a homeomorphism onto $\mathcal{X}_{\sigma}$.\label{homeo}
    \end{enumerate}
\noindent Condition \ref{cond.2} implies that the indexing set $\mathcal{P}_{\mathcal{X}}$ has a poset structure, given by $\tau$ $\leq$ $\sigma$ iff $\mathcal{X}_{\tau}$ $\subseteq$ $\overline{\mathcal{X}_\sigma}$, and we say that $\tau$ \textit{bounds} $\sigma$. This is known as the \textit{face poset} of $\mathcal{X}$. The regularity condition \ref{homeo} implies that all of the topological information about $\mathcal{X}$ is encoded in the poset structure of $\mathcal{P}_{\mathcal{X}}$. A regular cell complex can then be identified with its face poset. Thus, we will indicate the cell $\mathcal{X}_{\sigma}$ with its corresponding face poset element $\sigma$. The dimension $\textrm{dim}(\sigma)$ of a cell $\sigma$ is $k$; we call it a $k-$cell and denote it with $\sigma^k$ to make this explicit. 
A planar graph is a particular case of a regular cell complex of order 1, containing only cells of order 0 (nodes) and 1 (edges). An example of a cell complex of order 2 is a graph with order 2 cells being some of its induced cycles that we refer to as \textit{polygons}.

\noindent \textbf{Boundary.} The boundary \cite{sardellitti2022cellsp} of a $k$-cell $\sigma^k$ is the union of all $(k-1)$-cells bounding $\sigma^k$.  The dimension or order of a cell complex is the largest dimension of any of its cells, and we denote an order $K$ regular cell complex with $\mathcal{X}^K$. 

\noindent\textbf{Lower and upper neighborhoods} Fixed a cell dimension $k$, two neighborhoods among $k$-cells can be defined. In particular, we say that two $k-$cells are lower neighbors if they share a common face of order $k-1$ and upper neighbors if both are faces of a cell of order $k+1$. Thus, for example, two edges are lower adjacent if they share a common vertex, whereas they are upper adjacent if they are faces of a common polygon. We denote the upper and lower neighborhoods of cell $\sigma^k_i$ with $\mathcal{N}_u(\sigma^k_i)$ and $\mathcal{N}_d(\sigma^k_i)$, respectively.

\noindent \textbf{Topological Signals.} Let us denote the set of $k$-cells in $\mathcal{X}^{K}$ as ${\cal C}_{k} := \{\sigma_i^{k}: \sigma_i^{k} \in \mathcal{X}^{K}\} $, with $|{\cal C}_{k}| = N_k$.
        A $k$-topological signal over a regular cell complex $\mathcal{X}^K$ is defined as a collection of mappings from the set of all $k$-cells contained in the complex to real numbers:
        \begin{equation}\label{signals}
            \mathbf{x}_k = [x_k(\sigma_1^k),\dots, x_k(\sigma_{N_k}^k)]^T, \quad k=1,\ldots,K,
        \end{equation}
        where $\x_{k}: {\cal C}_{k} \rightarrow \mathbb{R}$.

\noindent \textbf{Cell Complex FIR Filters.} By directly generalizing the notion of graph shift operator \cite{ortega2018graph} to the cell complex domain, we introduce two real symmetric matrices $\S_k^{(d)}=\{s^{(d)}_{ij}\}_{i,j =1}^{N_k} \in \mathbb{R}^{N_k\times N_k}$ and  $\S_k^{(u)}=\{s^{(u)}_{ij}\}_{i,j =1}^{N_k}  \in \mathbb{R}^{N_k\times N_k}$ encoding upper and lower connectivity of the complex, i.e., such that $[\S_k^{(d)}]_{ij}=0$ if $\sigma_j^k\notin \mathcal{N}_{d}(\sigma_i^k)$, and $[\S_k^{(u)}]_{ij}=0$ if $\sigma_j^k\notin \mathcal{N}_{u}(\sigma_i^k)$. Multiplying a topological signal $\mathbf{x}_k$ by $\S_k^{(d)}$ or $\S_k^{(u)}$ performs a local (distributed) shift operation that replaces a signal value at each $k$-cell with the linear combination of the signal values over the lower and upper neighborhoods of the $k$-cell, respectively. Then, based on this local shift operation, cell complex finite impulse response (FIR) filters can be defined as \cite{barbarossa2020topological,yang2022simplicial,SPCC_hajij}:
    \begin{align}\label{topological_filter}
        \mathbf{y}_k &= \sum_{p=0}^P w_{p}^{(d)}(\S_k^{(d)})^p\x_k + \sum_{p=0}^P w_{p}^{(u)}(\S_k^{(u)})^p\x_k
    \end{align}
where $w_{p}^{(d)}$ and $w_{p}^{(u)}$ are the filter weights and $P$ is the filter length. 

\noindent\textbf{Cell Complex Convolutional Neural Networks.} The composition of a pointwise nonlinearity $\gamma(\cdot)$ and banks of filters as in \eqref{topological_filter} gives rise to cell complex convolutional neural networks \cite{hajij2023topological}, an instance of topological neural networks. In particular, let us assume that $F_{in}$ topological signals $\{\x_{k,f}\}_{f=1}^{F_{in}}$ are given as input to a layer of a cell complex convolutional neural network. The $F_{out}$ output signals $\{\y_{k,g}\}_{k,g=1}^{F_{out}}$ are computed as:
\begin{align}\label{TNN_not_mat}
\y_{k,g} = \gamma &\Bigg( 
\sum_{f=1}^{F_{in}} \sum_{p=0}^{P}\left[w_{p,f,g}^{(u)}\Big(\S_k^{(u)}\Big)^p +w_{p,f,g}^{(d)}\Big(\S_{k}^{(d)}\Big)^p\right]\x_{k,f}\Bigg )
\end{align}
for $g=1,...,F_{out}$. The filter weights are learnable parameters.  A cell complex convolutional neural network of depth $L$  is built as the stack of $L$ layers defined as in (\ref{TNN_not_mat}). For the sake of exposition, but without loss of generality, we will focus on the processing of edge signals. Thus, we will denote $\sigma^1$ with $\sigma$, $\x_1$ as $\x$, ${\cal C}_{1}$ as ${\cal C}$, $N_1$ as $N$, $\S^{(u)}_1$ and $\S^{(d)}_1$ as $\S^{(u)}$ and $\S^{(d)}$, respectively.

\begin{figure*}[h]
  \centering
  \includegraphics[width=.95\textwidth]{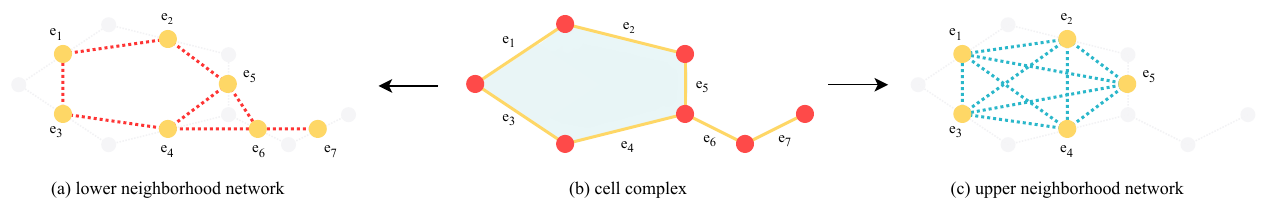} 
  \caption{Upper (a) and lower (c) communication networks induced by the cell complex built on the input network (b).}
  \label{fig::complex2networks} 
\end{figure*}

\section{Topological Neural Networks \\ over the Air }
\label{sec:format}

We consider a scenario in which data defined over the edges of the complex are collected by sensors, which can wirelessly exchange information according to two distinct communication topologies encoded by the lower and upper connectivity induced by the complex structure. In particular, given the cell complex structure, we assume the design of proper resource allocation strategies (e.g., power control, analog beamforming, etc.) that enable communication over the two distinct topologies. An example of upper and lower communication topologies induced by the relational complex structure is illustrated in Fig. \ref{fig::complex2networks}. Then, the signal value $[\x]_i$ of the $i$-th edge is transmitted in an uncoded/analog manner over two distinct time slots, implementing an over-the-air aggregation of cell information over the upper and lower neighborhoods, respectively. Assuming slow-fading channels $h^{(u)}_{ij}$ and $h^{(d)}_{ij}$ over both upper and lower neighborhoods, the signal value of cell $\sigma_i$ after the first communication and aggregation round is updated as: 
\begin{align}
        [\x^{(d,1)}]_i & = \Bigg (\sum_{\sigma_j\in\mathcal N_d(\sigma_i)} h_{ij}^{(d,1)}[\x]_j + n_{i}^{(d,1)} \Bigg) \quad \forall \ \sigma_i \in \mathcal C, \label{shifted_signal_down}\\
        [\x^{(u,1)}]_i & = \Bigg (\sum_{\sigma_j\in\mathcal N_u(\sigma_i)} h_{ij}^{(u,1)}[\x]_j + n_{i}^{(u,1)} \Bigg) \quad \forall \ \sigma_i \in \mathcal C, \label{shifted_signal_up}
\end{align}
where $n_i^{(d,1)}$ and $n_i^{(u,1)}$ are additive white Gaussian noise (AWGN) components. Channel fading effects are assumed to be independent across communication links, constant within a single communication round, and i.i.d. across successive transmissions. Eqs. \eqref{shifted_signal_down}-\eqref{shifted_signal_up} represent two noisy graph shift operations  of signal $\x$ applied with two different graphs and shift operators. In matrix form, we get:
\begin{align}
    &\x^{(d,1)} = \S^{(d,1)}_{\text{air}}\x+ \n^{(d,1)}\label{AirTSO_d}\\
      &\x^{(u,1)} = \S^{(u,1)}_{\text{air}}\x+ \n^{(u,1)}\label{AirTSO_u}  
\end{align}
where $[\n^{(d,1)}]_i=n_{i}^{(d,1)}$, $[\n^{(u,1)}]_i=n_{i}^{(u,1)}$, and the shift operators collect the channel gains as:
    \begin{align} 
    [\S^{(d,1)}_{\text{air}}]_{ij} &= h_{ij}^{(d,1)}\quad \forall \sigma_i\in \mathcal C, \;\;\forall \sigma_j\in\mathcal N_d(\sigma_i), \label{down_ota_shif}\\    
    [\S^{(u,1)}_{\text{air}}]_{ij} &= h_{ij}^{(u,1)}\quad \forall  \sigma_i\in \mathcal C,\;\; \forall \sigma_j\in\mathcal N_u(\sigma_i) \label{up_ota_shif}.
    \end{align}
Different from its ideal counterpart, the signal shifting in (\ref{AirTSO_d})-(\ref{AirTSO_u}) depends not only on the cell-complex topology, but also on the communication
channels, and we refer to the latter as the topological shift operation over the air (AirTSO). Specifically, an
AirTSO shifts $\x$ through wireless communication channels and obtains the one-shifted signals in (\ref{AirTSO_d})-(\ref{AirTSO_u}) that aggregates 1-hop neighborhood information over lower and upper connectivities, respectively. Recursively shifting $p$ times the signal $\x$, we obtain:
\begin{align}
\x^{(d,p)}
& =\  \S^{(d,p)}_{\text{air}}\left(\S^{(d,P-1)}_{\text{air}}(... + \n^{(d,p-2)})+\n^{(d,p-1)}\right)+\n^{(d,p)} \nonumber\\
\hspace{-.2cm}&=\ \prod_{\rho=1}^p \S^{(d,\rho)}_{\text{air}} \x + \sum_{i=1}^{p-1} \prod_{\rho=i+1}^p \S^{(d,\rho)}_{\text{air}}\n^{(d,i)} + \n^{(d,\rho)} \label{p_shift_d} \\
\x^{(u,p)}
& =\  \S^{(u,p)}_{\text{air}}\left(\S^{(u,p-1)}_{\text{air}}(... + \n^{(u,p-2)})+\n^{(u,p-1)}\right)+\n^{(u,p)}\nonumber\\
\hspace{-.2cm}&=\  \prod_{\rho=1}^p \S^{(u,\rho)}_{\text{air}} \x + \sum_{i=1}^{p-1} \prod_{\rho=i+1}^p \S^{(u,\rho)}_{\text{air}}\n^{(u,i)} + \n^{(u,\rho)} \label{p_shift_u} 
\end{align}
The multi-shifted signals in (\ref{p_shift_d})-(\ref{p_shift_u}) access farther nodes and aggregate the $p$-hop lower and upper neighborhood information, respectively, with communication noise.

\noindent \textbf{Topological Filter Over-the-Air (AirTF).} An AirTF is a linear combination of multi-shifted topological signals over the air. Given two sequences of shifted signals over lower and upper neighborhoods, i.e., $\{\x,\x^{(d,1)},\ldots,\x^{(d,P)}\}$ and $\{\x, \x^{(u,1)},\ldots,\x^{(u,P)}\}$ as in (\ref{p_shift_d})-(\ref{p_shift_u}), the AirTF can be written as:
\begin{align}\label{AirTF}
\mathbf{y}
=&\, \T_{\text{air}}(\S^{(d)},\S^{(u)}) \x =
\sum_{p=0}^{P} w_{p}^{(d)}\x^{(d,p)} + \sum_{p=0}^{P} w_{p}^{(u)}\x^{(u,p)},\nonumber\\
=& \, \left(\sum_{p=0}^{P} w_{p}^{(d)} \prod_{\rho=1}^p \S^{(d,\rho)}_{\text{air}}+\sum_{p=0}^{P} w_{p}^{(u)} \prod_{\rho=1}^p \S^{(u,\rho)}_{\text{air}}\right)\x + \nonumber\\
&\;+ \sum_{p=0}^{P} w_{p}^{(d)}\left( \sum_{i=1}^{p-1} \prod_{\rho=i+1}^p \S^{(d,\rho)}_{\text{air}}\n^{(d,i)} + \n^{(d,\rho)}   \right) + \nonumber\\
&\;+ \sum_{p=0}^{P} w_{p}^{(u)}\left( \sum_{i=1}^{p-1} \prod_{\rho=i+1}^p \S^{(u,\rho)}_{\text{air}}\n^{(u,i)} + \n^{(u,\rho)}   \right) 
\end{align}
where $\{w_{p}^{(d)}\}_{p=1}^P$ and $\{w_{p}^{(u)}\}_{p=1}^P$ are the set of filter weights, and P is the filter length. The AirTF is a shift-and-sum operator that extends topological convolution to wireless communication channels. It aggregates the neighborhood information up to a radius of $P$ and accounts for channel fading and Gaussian noise when shifting signals over the underlying cell complex topology. AirTF is intrinsically decentralized and designed to aggregate the information from the $P$-hop in both lower and upper neighborhoods. With perfect communication, i.e., $h_{ij}^{(d,p)}=h_{ij}^{(u,p)}=1$, and $n_i^{(d,p)}=n_i^{(u,p)}=0$, for all $i,p$ in (\ref{p_shift_d})-(\ref{p_shift_u}), AirTF simplifies to a standard cell complex filter as in \eqref{topological_filter}, and it represents also a generalization of the graph filters over-the-air proposed in \cite{airgnns}. 

\noindent \textbf{Topological Neural Networks Over-the-Air.} We are now able to introduce the topological neural network over-the-air (AirTNN) architecture. Following the same paradigm as in \eqref{TNN_not_mat}, we define an AirTNN layer as the composition of two main stages: i) a bank of topological filters over-the-air $\{\T_{\text{air}}^{fg}\}_{fg}$ as in (\ref{AirTF}), and ii) a pointwise nonlinearity $\gamma(\cdot)$. Then, assuming that $F_{in}$ topological signals $\{\x_f\}_{f}$ are given as input to a layer of AirTNN, the $F_{out}$ output signals $\{\y_g\}_g$ are computed as:
\begin{align}\label{AirTNNx}
\y_g &= \gamma\left(\sum_{f=1}^{F_{in}} \T_{\text{air}}^{fg}(\S^{(d)},\S^{(u)})\x_f  \right) \nonumber\\
&=\gamma \left( 
\sum_{f=1}^{F_{in}} \sum_{p=0}^{P}w_{p,f,g}^{(d)}\x_{f}^{(d,p)} + \sum_{f=1}^{F_{in}} \sum_{p=0}^{P}w_{p,f,g}^{(u)}\x_{f}^{(u,p)} \right)
\end{align}
for $g=1,...,F_{out}$, where $\x_f^{(d,p)}$ and $\x_f^{(u,p)}$ are the multi-shifted topological signals shifted over-the-air over lower and upper neighborhoods, respectively, as defined in (\ref{p_shift_d})-(\ref{p_shift_u}). Finally, the layer in (\ref{AirTNNx}) can be recast in compact matrix form as: 
\begin{align}\label{AirTNN}
    \mathbf{Y} = \gamma \Bigg ( \sum_{p=0}^{P} \X^{(u,p)}\W_p^{(u)} + \sum_{p=0}^{P} \X^{(d,p)}\W_{p}^{(d)} \Bigg ),
\end{align}
where $\X^{(u,p)}=\{\x^{(u,p)}_f\}_{f}\in \mathbb{R}^{N\times F_{in}}$ and $\X^{(d,p)}=\{\x^{(d,p)}_f\}_{f}\in \mathbb{R}^{N\times F_{in}}$ collect the multi-shifted topological signals, $\mathbf{Y} = \{\mathbf{y}_g \}_g$ collects the output signals, and  $\W^{(d)}={\{w^{(d)}_{k,f,g}\}}_{f,g}\in \mathbb{R}^{F_{in}\times F_{out}}$ and $\W^{(u)}={\{w^{(u)}_{k,f,g}\}}_{f,g}\in \mathbb{R}^{F_{in}\times F_{out}}$ collect the learnable filters weights. Therefore, an AirTNN of depth $L$ is built as the stack of $L$ layers as in (\ref{AirTNN}).

\noindent\textbf{Training of AirTNNs.} We train AirTNNs following the approach from \cite{airgnns}. We employ the usual gradient-based backpropagation optimizers (e.g., SGD, ADAM, etc.) but incorporating the randomness of the communication channels in the training. In particular, per each training step, we sample channel coefficients and AWGN to build the shift operators in \eqref{down_ota_shif}-\eqref{up_ota_shif} and the noise vectors in \eqref{shifted_signal_down}-\eqref{shifted_signal_up}. In this way, AirTNNs are able to learn filter weights that are robust to channel and noise conditions, as we will show in the sequel.

\section{Applications and Numerical Results}
\label{sec:pagestyle}

Our framework can be applied in scenarios where it is of interest to learn from topological data (e.g., defined over edges and/or polygons of a cell complex), which are collected by sensors that communicate over wireless channels to enable distributed processing. Several examples can be found in applications related to the monitoring and control of critical network infrastructure, such as traffic, hydraulic, or communication networks,
where flow measurements (i.e., edge signals, $k=1$ in \eqref{signals}) play a crucial role. In this context, AirTNNs jointly provide a principled learning method and a robust communication scheme, which relies on two different communication networks designed to let sensors exchange information between lower and upper neighborhoods induced by the cell complex structure. On the one hand, this feature provides a principled processing technique for flow signals due to the fact that AirTNNs are based on arguments from TSP \cite{barbarossa2020topological, giusti2021san}. On the other hand, the partial redundancy of communication links in the two (upper and lower) networks simplifies network design and makes communication more robust and resilient to failures. In the sequel, we test AirTNNs on a source localization task from flow data, which is interesting for anomaly detection. \footnote{\href{https://github.com/SimoneFiorellino/AirTNN.git}{https://github.com/SimoneFiorellino/AirTNN.git}}

\begin{figure}[t]
    \centering    \includegraphics[width=.8\columnwidth]{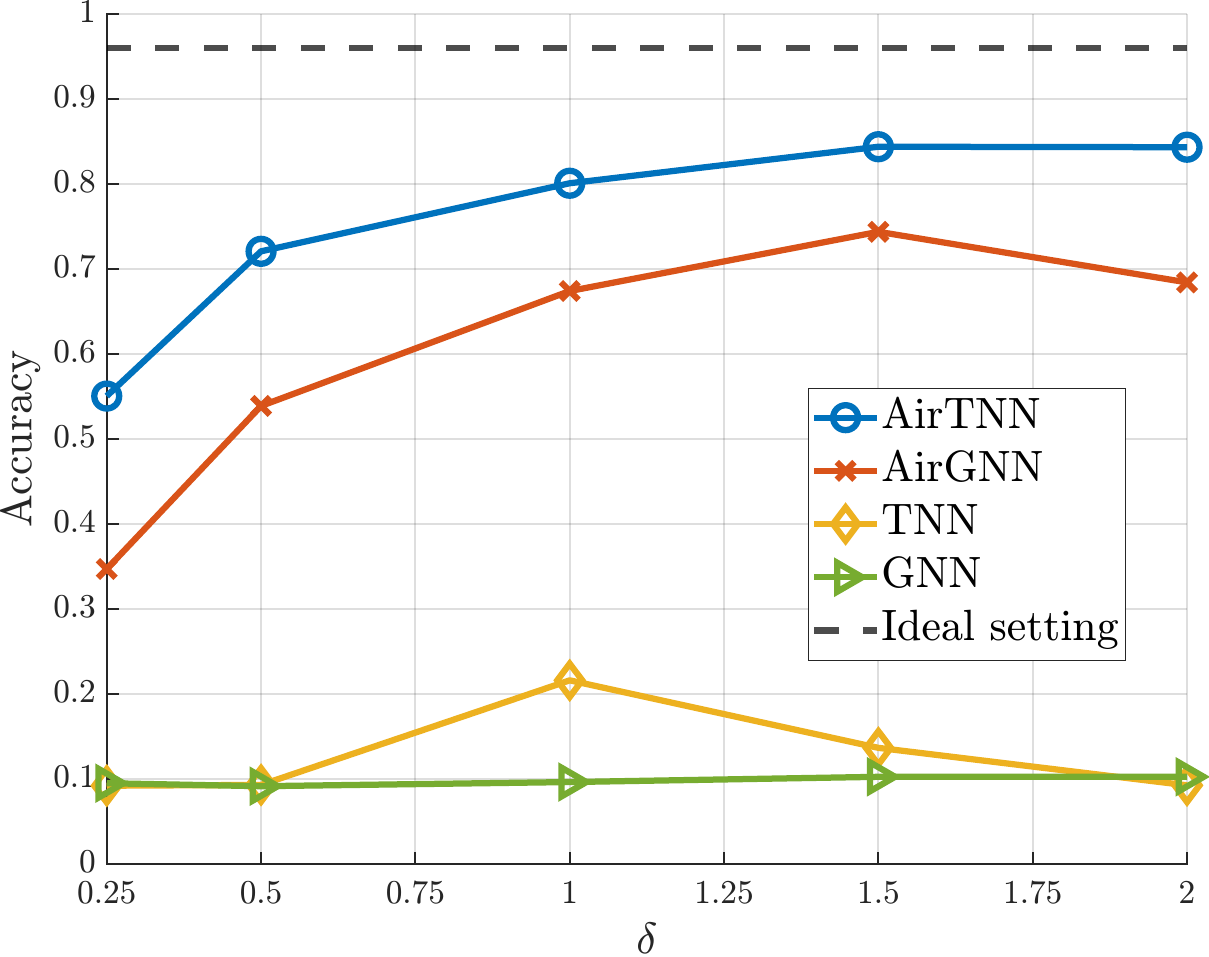}
    \caption{Accuracy versus $\delta$, with a fixed SNR of 20 dB.}
    \label{fig::res_delta}
\end{figure}

\noindent \textbf{Dataset design.} We build a network using a Stochastic Block Model (SBM) \cite{holland1983stochastic}; the resulting graph is composed of 70 nodes uniformly divided into 10 communities, with intra- and inter-community edge probabilities of 0.9 and 0.01, respectively. The intra-community edges from the original graph are grouped to form 10 edge communities in an equivalent representation. Additionally, we consider an 11th partition of edges composed of inter-community edges. The graph is then extended to a cell complex by considering the minimum collection of polygons (i.e., a cycle basis) to cover the overall complex \cite{paton1969algorithm}. We generate synthetic edge signals according to the model $\x_1 = \B_1^T \x_0 + \B_2 \x_2$, where $\x_0$ denotes the nodes signal, $\x_2$ the polygons signal, $\B_1$ and $\B_2$ are the edge-nodes and edge-polygons incidence matrices, respectively \cite{barbarossa2020topological}. The nodes and polygons signals $\x_0$ and $\x_2$ are sampled from a zero-mean Gaussian distribution with variance ${1}/{N_1}$.  The aim of this experiment is source localization at the edge level. Specifically, the objective is to identify the $c$-th community from which $\eta$ diffused spikes have originated. Thus, having $\{s_i\}_{i=1}^{N_1}$ potential sources, we define $\mathcal{D}_c$ as the set of the sources belonging to the $c$-th cluster. Each spike is modeled as a Kronecker delta with intensity $\alpha \sim \mathcal{N}(0, \psi)$, with $\psi\in\mathbb R$. Then, the spikes added to the cluster $c$ are represented by $\tilde \delta_c = \sum_{\forall s_i\in\mathcal{I}_c} ( \alpha \delta_i )$, with $\mathcal{I}_c$ being the collection of $\eta$ randomly selected sources belonging to $\mathcal{D}_c$, chosen without replacement. Then, we assume the flow signal with spikes is diffused over the cell complex topology according to $\x_{c,\tau} = \overline{\S}^\tau (\x_1 + \tilde\delta_c) +\n$, with $\overline{\S} = {\S}/{\lambda_{\text{max}}(\S)}$, $\n$ as additive white Gaussian noise with Signal-to-Noise Ratio (SNR) set at 40 dB, and $\tau$ the diffusion order of the signal. 
For each sample, the cluster $c$ and the diffusion parameter $\tau$ are chosen at random. The shift operator, $\S$, is selected as the lower  shift operator $\S^{(d)}_k$, respecting a possible diffusion over a physical/spatial undirected flow network.


\begin{figure}[t]
    \centering
\includegraphics[width=.8\columnwidth]{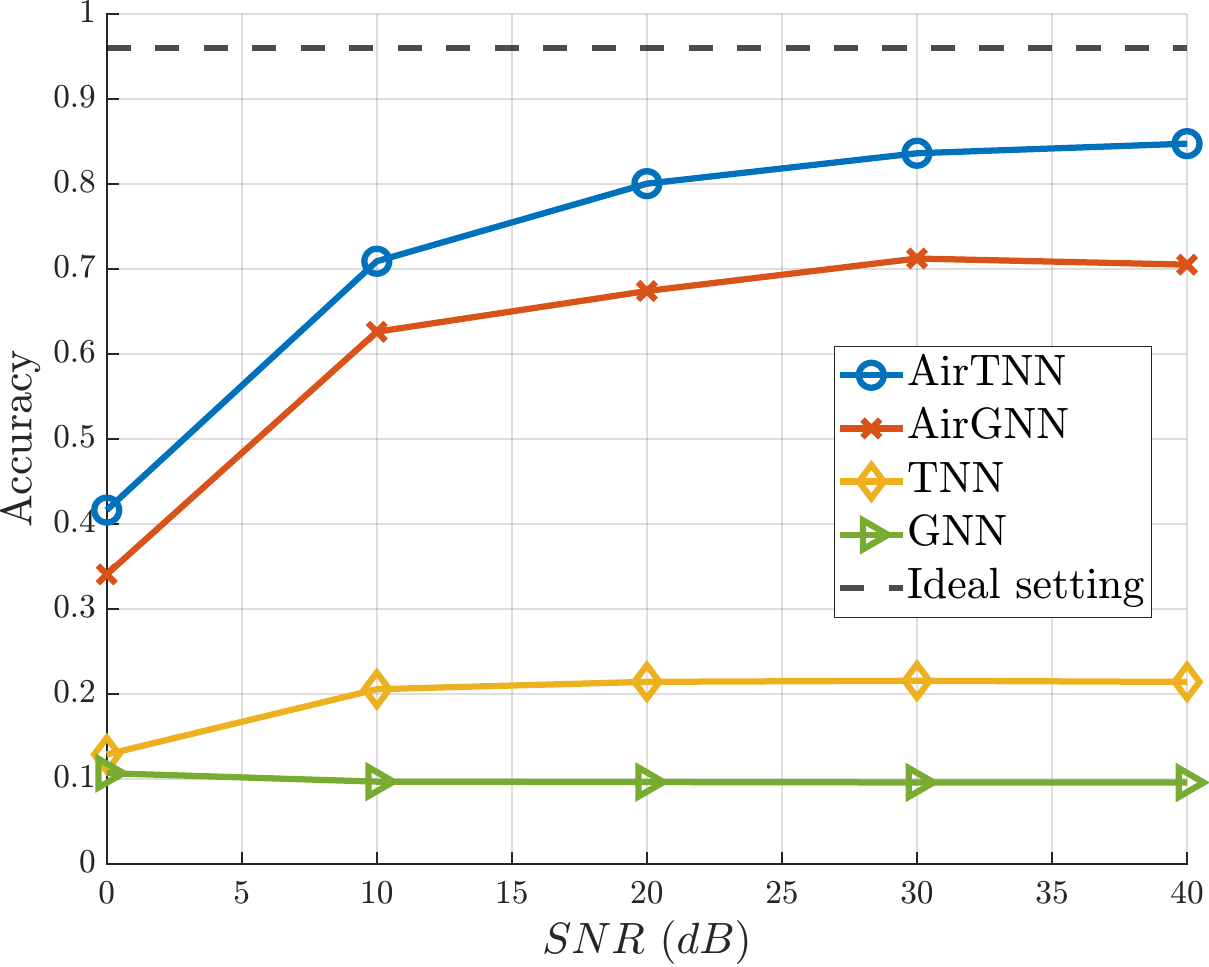}
    \caption{Accuracy versus SNR, with a fixed $\delta$ equal to 1.}
    \label{fig::res_snrdb}
\end{figure}

\noindent \textbf{Experimental setup and discussion.} The experiment compares models in the graph and cell complex domains. In the graph domain, communication occurs between edges connected through nodes, i.e., over the lower communication network (Fig. \ref{fig::complex2networks}a). 
We compared AirTNN with five distinct models, each structured with two layers followed by a readout component leading to a final feed-forward neural network. The considered models include: (i) TNN and (ii) graph neural networks (GNN), which both operate without channel fading and noise (i.e., ideal communication) during both training and testing, serving as baseline models; 
(iii) TNN and (iv) GNN, which are affected by channel fading and noise, but only during testing;  and (v) AirGNN from \cite{airgnns}. 
Channel fading is Rayleigh distributed with scale parameter $\delta$, and communication noise is zero-mean white Gaussian, with variance depending on the chosen SNR. In Fig. \ref{fig::res_delta} and Fig.  \ref{fig::res_snrdb}, we present the results of two experiments. Specifically, in Fig. \ref{fig::res_delta}, we illustrate the accuracy of the considered models versus the channel parameter $\delta$, while keeping the SNR fixed to 20 dB. Then, in Fig. \ref{fig::res_snrdb}, we report the behavior of the accuracy of the models versus the SNR, considering $\delta=1$. As we can notice from Figs. \ref{fig::res_delta} and \ref{fig::res_snrdb}, AirTNN consistently demonstrates enhanced robustness against disturbances introduced by wireless communication. This resilience can be attributed to the AirTNN design, which is specifically built to process information defined over edge flows, exploiting two distinct neighborhoods that experience different channel and noise conditions at each step. This feature sets AirTNN apart from AirGNN and enables enhanced performance, as we can see from Figs. \ref{fig::res_delta} and \ref{fig::res_snrdb}. As expected, neither GNN nor TNN can accomplish the task, since they do not consider communication impairments in the training process.

\vspace{-.3cm}
\section{Conclusions}
\vspace{-.2cm}
\label{sec:typestyle}

In this paper, we have proposed AirTNN, a novel architecture that performs over-the-air distributed processing of data defined over regular cell complexes, integrating the wireless communication model into its architecture. Specifically, during training and inference, the proposed method considers channel impairments such as fading and noise in the topological convolutional filtering operation, which takes place over different signal orders and neighborhoods. As a result, AirTNNs are multi-layered architectures built by cascading the novel cell complex filters over the air and pointwise nonlinearities. Numerical experiments illustrate that the proposed architecture outperform both AirGNNs and baseline GNNs and TNNs.

\clearpage
\balance
\bibliographystyle{IEEEbib}
\bibliography{main}

\end{document}